\documentclass[11pt, a4paper]{article}
\pdfoutput=1
\usepackage{jcappub}

\usepackage{booktabs}
\usepackage{multirow}
\usepackage{amssymb}
\usepackage[export]{adjustbox}
\usepackage{floatrow}
\newfloatcommand{capbtabbox}{table}[][3.5in]
\newfloatcommand{capbfigbox}{figure}[][2.3in]
\usepackage{blindtext}
\usepackage{amsmath}
\usepackage{booktabs}
\usepackage{aas_macros}

\newcommand{\like}{\ensuremath{\mathcal{L}}}

\def\lsim{\mathrel{\raise.3ex\hbox{$<$\kern-.75em\lower1ex\hbox{$\sim$}}}}
\def\gsim{\mathrel{\raise.3ex\hbox{$>$\kern-.75em\lower1ex\hbox{$\sim$}}}}

\begin{document}

\hspace*{110mm}{\large \tt FERMILAB-PUB-16-128-A}

\vskip 0.2in

\title{Radio Galaxies Dominate the High-Energy Diffuse Gamma-Ray Background}

\author[a,b,c]{Dan Hooper}
\emailAdd{dhooper@fnal.gov}
\author[d]{Tim Linden}
\emailAdd{linden.70@osu.edu}
\author[e,a]{and Alejandro Lopez}
\emailAdd{aolopez@umich.edu}

\affiliation[a]{Fermi National Accelerator Laboratory, Center for Particle
Astrophysics, Batavia, IL 60510}
\affiliation[b]{University of Chicago, Department of Astronomy and Astrophysics, Chicago, IL 60637}
\affiliation[c]{University of Chicago, Kavli Institute for Cosmological Physics, Chicago, IL 60637}
\affiliation[d]{Ohio State University, Center for Cosmology and AstroParticle Physcis (CCAPP), Columbus, OH  43210}
\affiliation[e]{Michigan Center for Theoretical Physics, Department of Physics, University of Michigan, Ann Arbor, MI  48109}

\abstract{It has been suggested that unresolved radio galaxies and radio quasars (sometimes referred to as misaligned active galactic nuclei) could be responsible for a significant fraction of the observed diffuse gamma-ray background. In this study, we use the latest data from the Fermi Gamma-Ray Space Telescope to characterize the gamma-ray emission from a sample of 51 radio galaxies. In addition to those sources that had previously been detected using Fermi data, we report here the first statistically significant detection of gamma-ray emission from the radio galaxies 3C 212, 3C 411, and B3 0309+411B. Combining this information with the radio fluxes, radio luminosity function, and redshift distribution of this source class, we find that radio galaxies dominate the diffuse gamma-ray background, generating $77.2^{+25.4}_{-9.4}\%$ of this emission at energies above $\sim$1 GeV. We discuss the implications of this result and point out that it provides support for scenarios in which IceCube's high-energy astrophysical neutrinos also originate from the same population of radio galaxies.}

\maketitle

\section{Introduction}

Over the decades since its discovery~\cite{1978ApJ...222..833F,Sreekumar:1997un,Abdo:2010nz,Ackermann:2014usa}, significant progress has been made in our understanding of the origin of the diffuse extragalactic gamma-ray background. Although it has long been speculated that this emission is likely to originate from a large number of unresolved sources, the nature of those sources has only recently become more apparent. In particular, it is now known that this background receives non-negligible contributions from several classes of sources, including blazars (flat-spectrum radio quasars and BL Lac objects)~\cite{Cuoco:2012yf,Harding:2012gk,Ajello:2011zi,Ajello:2013lka}, radio galaxies~\cite{Inoue:2011bm,Fields:2010bw}, and star-forming galaxies~\cite{Tamborra:2014xia,Ackermann:2012vca}, along with smaller contributions from galaxy clusters~\cite{Zandanel:2014pva}, millisecond pulsars~\cite{Calore:2014oga,Hooper:2013nhl}, and propagating ultra-high energy cosmic rays~\cite{Taylor:2015rla,Ahlers:2011sd}. At this time, it appears plausible that a combination of these source classes could account for the observed intensity and spectrum of the diffuse gamma-ray background~\cite{DiMauro:2016cbj,DiMauro:2015tfa,Ajello:2015mfa,Cholis:2013ena,Cavadini:2011ig,Siegal-Gaskins:2013tga}. It is also possible that annihilating dark matter particles may contribute~\cite{Ackermann:2015tah,DiMauro:2015tfa,Ajello:2015mfa,Cholis:2013ena}. In addition to spectral measurements, further clarity has resulted from Fermi's measurement of the small-scale anisotropy of the diffuse gamma-ray background~\cite{Ackermann:2012uf}, as well as the cross-correlations of this background with multi-wavelength data~\cite{Xia:2015wka,Cuoco:2015rfa,Shirasaki:2014noa,Shirasaki:2015nqp}. Despite all of this progress, however, there are still many questions pertaining to the origin of the diffuse gamma-ray background that remain unanswered (for a review, see Ref.~\cite{Fornasa:2015qua}).

In this paper, we focus on the contribution from radio galaxies and radio quasars (which we will collectively refer to simply as ``radio galaxies''\footnote{Such sources are also sometimes referred to as ``misaligned active galactic nuclei''.}) to the diffuse gamma-ray background. Within the context of the unified model of active galactic nuclei (AGN)~\cite{Urry:1995mg}, blazars are the subset of AGN whose jets are directed within approximately $14^{\circ}$ of Earth. In contrast, AGN with jets oriented in other directions appear fainter, and are categorized as radio galaxies. Radio galaxies are further classified as either Fanaroff-Riley Type I or Type II galaxies (FRI or FRII, respectively), according to their morphological characteristics. FRIs and FRIIs are generally interpreted as the misaligned counterparts of BL Lacs and flat spectrum radio quasars, respectively. Although blazars are individually much more luminous than radio galaxies (explaining the much greater number of blazars that have been resolved by Fermi), radio galaxies are much more numerous, making it possible for them to collectively contribute significantly to the diffuse gamma-ray background~\cite{Inoue:2011bm}.

Here, we revisit the contribution to the diffuse gamma-ray background from unresolved radio galaxies. In doing so, we utilize the most recent data set from the Fermi Gamma-Ray Space Telescope to refine the previously observed correlation between the radio and gamma-ray emission detected from radio galaxies. We then make use of this correlation, along with the radio luminosity function and redshift distribution of the population, to calculate the total contribution to the diffuse gamma-ray background. We find that unresolved radio galaxies dominate the diffuse gamma-ray background at energies above 1 GeV, accounting for $77.2^{+25.4}_{-9.4}\%$ of the observed emission. The results presented here also support the possibility that radio galaxies may be responsible for the flux of astrophysical neutrinos observed by IceCube.

\section{Gamma-Rays From Radio Galaxies}
\label{gamma}

The Fermi Gamma-Ray Space Telescope has detected statistically significant emission from a number of radio galaxies. In particular, the Third Fermi Gamma-Ray Source Catalog (the 3FGL)~\cite{TheFermi-LAT:2015hja} includes 14 sources associated with radio galaxies\footnote{Although there are actually 15 sources in the 3FGL associated with radio galaxies, this list includes both the core and lobes of Cen A, which we choose not to count independently.}. In this study, we make use of the Fermi data from the directions of the 51 radio galaxies listed in Tables~\ref{table1} and~\ref{table2}. This sample of sources is identical to that adopted in Ref.~\cite{DiMauro:2013xta}, and was chosen to include those radio galaxies with the highest radio core fluxes at 5 GHz~\cite{2011ApJ...740...29K,2005A&A...432..401G}, excluding those located near the Galactic Plane or that exhibit a high degree of variability. 

For each of these 51 sources, we have determined the intensity and spectrum of their gamma-ray emission, utilizing 85 months of Fermi-LAT data\footnote{MET Range: 239557417 --- 464084557}. We have restricted our analysis to events which pass the Pass 8 Source event selection criteria and which lie in the energy range of 0.1 to 100 GeV. We apply standard cuts to the data, removing events that were recorded at a zenith angle larger than 90$^\circ$, while the instrument is not in Survey mode, while the instrumental rocking angle exceeds 52$^\circ$, or while Fermi was passing through the South Atlantic Anomaly. Given the low luminosities of many radio galaxies, we place no constraints on the point spread function class and include both front and back converting events. 

For each radio galaxy in our sample, we divide the resulting data set into 15 logarithmic energy bins as well as 280$\times$280 angular bins spanning a 14$^\circ\times$14$^\circ$ region-of-interest centered on the position of the source (as determined from radio observations). We then utilize a two step process to fit the resulting normalization and spectrum of each radio galaxy. First, we fit all background components \emph{except} for the radio galaxy over the full sky and over the full energy range using a spectral model for each source. In this stage we include the full 3FGL point source catalog~\cite{TheFermi-LAT:2015hja} and the {\tt gll\_iem\_v06.fits} Fermi diffuse emission model, as recommended by the Fermi Collaboration for Pass 8 data. We also utilize the matching isotropic background model {\tt iso\_P8R2\_SOURCE\_V6\_v06.txt}. We use the standard Fermi-LAT algorithm to determine whether a given source component should be allowed to float freely, or be held fixed in the fit, and use the python implementation of the {\tt gtlike} tool, including the {\tt MINUIT} algorithm, to determine the best-fit normalization and spectrum of each emission component.


\begin{table}
\renewcommand{\arraystretch}{1.2}
\begin{tabular}{|c|c|c|c|c|c|}
\hline 
Galaxy Name(s) & 3FGL Name & Redshift & Flux (erg/cm$^2$/s) & $\Gamma$ & TS \tabularnewline
\hline 
\hline 
3C~120 & -- & $0.0330$ & $6.9 _{-1.7}^{+1.1}\times10^{-12}$ & $2.47\pm0.08$& 31.0\tabularnewline
\hline 
3C~212 & -- & $1.049$ & $2.2_{-0.7}^{+0.8}\times10^{-12}$ & $2.56\pm0.16$& 25.1\tabularnewline
\hline 
3C~411 & -- & $0.467$ & $3.5_{-1.0}^{+1.1}\times10^{-12}$ & $2.85\pm0.17$& 27.3 \tabularnewline
\hline 
B3~0309+411B & -- & $0.134$ & $2.5_{-0.8}^{+1.0}\times10^{-12}$ & $1.98\pm0.19$& 32.1 \tabularnewline
\hline
3C~78/NGC~1218 & J0308.6+0408 & $0.0287$ & $6.6_{-1.1}^{+1.2}\times10^{-12}$ & $1.99\pm0.08$& 176. \tabularnewline
\hline 
3C~111 & J0418.5+3813c & $0.0485$ & $1.4_{-0.1}^{+0.2}\times10^{-11}$ & $2.68\pm0.06$& 249. \tabularnewline
\hline 
Pictor~A & J0519.2-4542 & $0.0351$ & $3.9_{-0.7}^{+0.8}\times10^{-12}$ & $2.35\pm0.1$& 84.0 \tabularnewline
\hline 
PKS~0625-35 & J0627.0-3529 & $0.0546$ & $1.7_{-0.2}^{+0.2}\times10^{-11}$ & $1.84\pm0.04$&  916.  \tabularnewline
\hline 
3C~207 & J0840.8+1315 & $0.681$ & $4.6_{-0.8}^{+0.8}\times10^{-12}$ & $2.52\pm0.09$&  97.7  \tabularnewline
\hline 
3C~264 & J1145.1+1935 & $0.021$ & $3.4_{-0.8}^{+1.0}\times10^{-12}$ & $1.84\pm0.15$&  91.8 \tabularnewline
\hline 
3C~274/M 87& J1230.9+1224 & $0.0038$ & $1.7_{-0.2}^{+0.2}\times10^{-12}$ & $1.96\pm0.04$&     949. \tabularnewline
\hline 
Cen~A (core) & J1325.4-4301 & $0.0009$ & $4.6_{-0.1}^{+0.2}\times10^{-11}$ & $2.39\pm0.02$& 3270. \tabularnewline
\hline 
3C~303 & J1442.6+5156 & $0.141$ & $1.5_{-0.5}^{+0.8}\times10^{-12}$ & $2.21\pm0.03$& 34.0  \tabularnewline
\hline 
NGC~6251 & J1630.6+8232 & $0.0247$ & $1.2_{-0.1}^{+0.1}\times10^{-11}$ & $2.28\pm0.03$& 838. \tabularnewline
\hline 
3C~380 & J1829.6+4844 & $0.692$ & $1.6_{-0.1}^{+0.1}\times10^{-11}$ & $2.28\pm0.07$& 1050. \tabularnewline
\hline 
Cen~B & J1346.6-6027 & $0.0129$ & $2.1_{-0.2}^{+0.3}\times10^{-11}$ & $1.81\pm0.20$& 221. \tabularnewline
\hline 
\end{tabular}
\caption{Selected characteristics for the 16 radio galaxies in our sample that Fermi detects at a level of TS $>25$.  For the 12 of these sources that are listed in the most recent Fermi source catalog, we include their 3FGL names~\cite{TheFermi-LAT:2015hja}. For each source, we list the gamma-ray flux (integrated between 0.1 and 100 GeV) and spectral index, including the 1$\sigma$ uncertainties, assuming a power-law form, $dN_{\gamma}/dE_{\gamma} \propto E_{\gamma}^{-\Gamma}$.}
\label{table1}
\end{table}

After the background fitting is completed in the region surrounding each radio galaxy, we add a point source at the position of the source, and perform a full scan of the likelihood fit as a function of the flux \emph{independently in each energy bin}. We then fit the resulting likelihood distribution assuming a number of arbitrarily normalized power-law spectra,  ${dN_{\gamma}/dE_{\gamma} \propto E_{\gamma}^{-\Gamma}}$, allowing the spectral index, $\Gamma$, to float between 0 and 3.5. By determining the maximum improvement to the likelihood, we calculate the value of the test statistic (TS) for each radio galaxy, as well as the likelihood profiles for the gamma-ray flux and spectral index.

We note that this two step method for fitting the normalization and spectrum of gamma-ray point sources has been used in many studies (e.g.~Ref.~\cite{Drlica-Wagner:2015xua}), and it has been shown that the removal of low-significance gamma-ray sources from the background fit produces no change in the background parameters. However, this may not be strictly true for the case where bright 3FGL sources are removed from the initial iteration of the fit. As these sources also exist in the 3FGL catalog, we have directly checked the calculated normalization and spectrum of each source compared to the values determined in 3FGL, and found that they are consistent.


For six of the radio galaxies in our sample (For A, Cen A, 3C 66B, 3C 83.1B, PKS 2153-69, and 3C 120) there is a 3FGL source located within 0.5$^\circ$ that is not associated with the radio galaxy under consideration. Because our
analysis fits the gamma-ray background to the data before calculating
the gamma-ray flux from each radio galaxy, the existence of a nearby
source could systematically decrease the
gamma-ray flux measured from these six radio galaxies. In order to correct for
this effect, we produce a grid of
background fits to the gamma-ray data for each of these six systems, changing both the normalization
and the spectral index of the background source from
its default value. For each point in our background grid, we then
calculate the best-fit flux of the radio galaxy as specified
above, and then marginalize our results over all background models,
taking the best-fit background model for each choice of the radio
galaxy normalization and spectrum. For most of these systems, this procedure had only a marginal effect, and our results would not change qualitatively if we had not made this correction.

\begin{table}
\begin{tabular}{|c|c|c|c|}
\hline 
Galaxy Name & Redshift & Flux (ergs/cm$^2$/s) & TS  \tabularnewline
\hline 
\hline 
3C 109 & $0.306$ & $<2.1\times10^{-12}$ & 1.0 \tabularnewline
\hline 
3C 18 & $0.188$ & $<2.5\times10^{-12}$ & 6.4\tabularnewline
\hline 
3C 215 & $0.412$ & $<1.7\times10^{-12}$ & 2.5\tabularnewline
\hline 
3C 227 & $0.086$ & $<1.0\times10^{-12}$ & 0.0 \tabularnewline
\hline 
3C 245 & $1.029$ & $<3.1\times10^{-12}$ & 16.2 \tabularnewline
\hline 
3C 272.1 & $0.003$ & $<2.4\times10^{-12}$& 7.1 \tabularnewline
\hline 
3C 277.3 & $0.0857$ & $<2.0\times10^{-12}$&  7.1 \tabularnewline
\hline 
3C 288 & $0.246$ & $<1.6\times10^{-12}$ & 3.9 \tabularnewline
\hline 
3C 29 & $0.045$ & $<2.5\times10^{-12}$ & 5.2  \tabularnewline
\hline 
3C 293 & $0.045$ & $<1.7\times10^{-12}$&  10.3  \tabularnewline
\hline 
3C 296 & $0.024$ & $<1.4\times10^{-12}$ &  0.6  \tabularnewline
\hline 
3C 305 & $0.0414$ & $<1.2\times10^{-12}$& 2.3 \tabularnewline
\hline 
3C 31 & $0.017$ & $<1.5\times10^{-12}$ & 1.5 \tabularnewline
\hline 
3C 310 & $0.054$ & $<2.2\times10^{-12}$& 6.6 \tabularnewline
\hline 
3C 315 & $0.1083$ & $<1.7\times10^{-12}$& 2.2 \tabularnewline
\hline 
3C 338 & $0.030$ & $<7.5\times10^{-13}$& 0.0 \tabularnewline
\hline 
3C 346 & $0.162$ & $<3.8\times10^{-12}$ & 24.2 \tabularnewline
\hline 
3C 348 & $0.1540$ & $<2.4\times10^{-12}$& 4.0 \tabularnewline
\hline 
3C 382 & $0.058$ & $<2.2\times10^{-12}$ & 5.3 \tabularnewline
\hline 
3C 386 & $0.018$ & $<2.3\times10^{-12}$ & 0.9 \tabularnewline
\hline 
3C 390.3 & $0.056$ & $<2.3\times10^{-12}$& 15.0 \tabularnewline
\hline 
3C 424 & $0.1270$ & $<2.1\times10^{-12}$ & 2.5  \tabularnewline
\hline 
3C 433 & $0.102$ & $<1.5\times10^{-12}$ & 0.3  \tabularnewline
\hline 
3C 438 & $0.290$ & $<1.7\times10^{-12}$ & 0.4 \tabularnewline
\hline 
3C 442A & $0.027$ & $<2.6\times10^{-12}$ & 6.0 \tabularnewline
\hline 
3C 445 & $0.056$ & $<1.3\times10^{-12}$ & 0.1 \tabularnewline
\hline 
3C 449 & $0.017$ & $<1.7\times10^{-12}$ & 1.3 \tabularnewline
\hline 
3C 465 & $0.029$ & $<1.6\times10^{-12}$& 0.6\tabularnewline
\hline 
3C 66B & $0.022$ & $<2.4\times10^{-12}$ & 0.0 \tabularnewline
\hline 
3C 83.1B & $0.026$ & $<3.1\times10^{-12}$& 2.0 \tabularnewline
\hline 
3C 89 & $0.1386$ & $<1.4\times10^{-12}$ & 0.0  \tabularnewline
\hline 
4C 74.26 & $0.104$ & $<2.7\times10^{-12}$& 9.5 \tabularnewline
\hline 
DA 240 & $0.036$ & $<1.4\times10^{-12}$ &  1.5 \tabularnewline
\hline 
PKS 2153-69 & $0.028$ & $<2.4\times10^{-12}$& 1.3 \tabularnewline
\hline
For A & 0.00587  & $<4.2 \times10^{-12}$  &11.5 \tabularnewline
\hline 
\end{tabular}
\caption{Upper limits on the gamma-ray flux (integrated between 0.1 and 100 GeV) for those radio galaxies considered in our analysis which did not yield a statistically significant detections (TS $< 25$).}
\label{table2}
\end{table}

\begin{figure}
\includegraphics[keepaspectratio,width=0.95\textwidth]{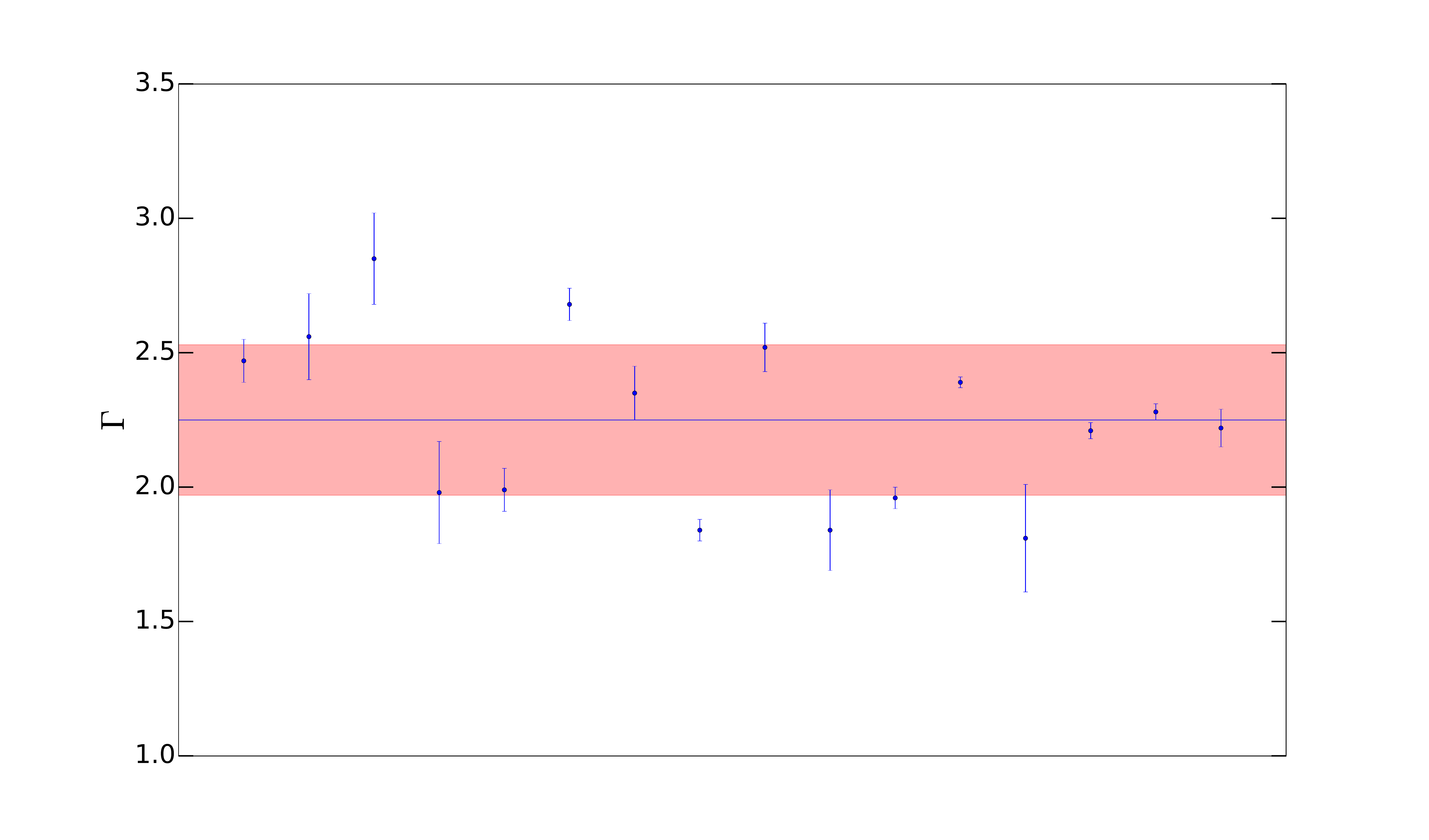}
\caption{The spectral index of the gamma-ray emission from each radio galaxy in our sample detected by Fermi with TS$>$25. The distribution of the spectral indices of this source population is well described by a Gaussian with a mean value of $\bar{\Gamma}=2.25$ and a width of $\sigma_{_\Gamma}=0.28$, as represented by the red band.}	
\label{index}		
\end{figure}

In Table~\ref{table1}, we list the gamma-ray flux (integrated between 0.1 and 100 GeV) and spectral index for each of the radio galaxies in our sample with gamma-ray emission detected at a level of TS $>$ 25\footnote{Note that although the sources included in Table~\ref{table1} largely overlap with the radio galaxies listed in the 3FGL, we have not included NGC 1275 or IC 310 in our study due to their high degree of radio variability. Also, we do not include 4C +39.12 or NGC 2484 due to the lack of 5 GHz flux measurements for these sources.}. Of these 16 sources, 12 are listed in the 3FGL catalog. In addition, we have identified significant gamma-ray emission from four other radio galaxies, including 3C 120, which was previously detected using Fermi data~\cite{2010ApJ...720..912A}, as well as 3C 212, 3C 411, and B3 0309+411B, which had not previously been reported as statistically significant gamma-ray sources (TS$\sim$10  detections of 3C 212 and 3C 411 were reported in Refs.~\cite{Liao:2015aua,Kataoka:2011hk}). The errors quoted in this table denote the 1$\sigma$ range for each quantity, as determined using the full 2D likelihood profile.  In Table~\ref{table2}, we list the (2$\sigma$) upper limits on the gamma-ray flux from each of the radio galaxies studied in our analysis that did not yield a statistically significant detection\footnote{We note that For A was associated with a fairly bright gamma-ray source in the 2FGL catalog~\cite{Fermi-LAT:2011iqa}, but not in the more recent 3FGL~\cite{TheFermi-LAT:2015hja}. Our analysis of the Fermi data in this region of the sky confirms that the observed gamma-ray emission associated with the source 3FGL J0322.5-3721 does not primarily originate from the direction of For A, and we place only an upper limit on the gamma-ray flux from this radio galaxy.}.

Of the 16 sources listed in Table~\ref{table1}, most of their fluxes are in good agreement with those reported by the Fermi Collaboration. The exception to this is the radio galaxy Cen A, for which the flux identified in our analysis ($4.6_{-0.1}^{+0.2}\times10^{-11}$ erg/cm$^2$/s) is equal to about 70\% of the value reported in the 3FGL. This is likely connected to the fact that Cen A is a spatially extended gamma-ray source, with an angular extent comparable to Fermi's point spread function. In any case, a 30\% shift in the flux of any one radio galaxy will have a negligible impact on the main results of this study. We also note that although most of the spectral indices listed in Table~\ref{table1} are in good agreement with those reported in the 3FGL, our value for Cen A's spectral index ($\Gamma=2.39 \pm 0.02$) is significantly softer than that listed in the 3FGL ($\Gamma=2.70\pm0.03$). Again we attribute this difference to challenges related to Cen A's spatial extension. To a lesser extent, our analysis also found somewhat softer spectral indices for Cen B and 3C 111, although only at the level of 2.3 and 1.1$\sigma$, respectively.

In Fig.~\ref{index}, we show the spectral indices of the gamma-ray emission from each of the radio galaxies detected by Fermi with TS $>$ 25. These spectral indices are largely concentrated around values between approximately 2.0 and 2.5, consistent with previous determinations~\cite{TheFermi-LAT:2015hja,DiMauro:2013xta}. To determine the underlying distribution of the spectral indices for this source population, we fit the likelihood function of each source, marginalized over the normalization of the flux, to a Gaussian, yielding values of $\Gamma_{i}$ and $\sigma_{\Gamma,i}$ for each radio galaxy. We then find the central value of the spectral index distribution, $\bar{\Gamma}$, for which the following function is minimized:
\begin{equation}
\chi^2 (\bar{\Gamma},\sigma_\Gamma ) =\sum_i \frac{\left(\bar{\Gamma} - \Gamma_i \right)^2 }{\sigma_{\Gamma,i}^2+\sigma_\Gamma^2}.
\end{equation}
The width of the underlying distribution, $\sigma_{\Gamma}$, is chosen such that $\chi^2(\bar{\Gamma},\sigma_\Gamma)$ is equal to the number of radio galaxies in the fit. From this procedure, we find that the spectral indices of this population of sources are well described by a Gaussian distribution centered at $\bar{\Gamma}=2.25$ and with a width of $\sigma_{_\Gamma}=0.28$ (shown as a red band in Fig.~\ref{index}). We further note that we have not identified any statistically significant correlations between the spectral index of the gamma-ray emission and other observed characteristics, such as radio or gamma-ray luminosities.


\section{The Gamma Ray-Radio Correlation}
\label{corsec}

If the gamma-ray emission observed from this collection of sources was our only information, it would be very difficult to make a reliable estimate for the total gamma-ray emission from all unresolved radio galaxies. To circumvent these limitations, one must also make use of the information provided by radio observations of this source population, namely the observed radio luminosity function and redshift distribution. The previously established correlation between the gamma-ray and radio emission from radio galaxies~\cite{Inoue:2011bm,DiMauro:2013xta}, which we will refine here, will allow us to calculate the expected contribution to the diffuse gamma-ray background from the global population of unresolved radio galaxies. 

In this study, we revisit the empirical correlation between the radio and gamma-ray luminosities from radio galaxies, making use of the full data set currently provided by Fermi (for the 51 sources listed in Tables~\ref{table1} and~\ref{table2}). As in previous studies~\cite{Inoue:2011bm,DiMauro:2013xta}, we adopt the following linear relationship for a galaxy's radio and gamma-ray luminosities:
\begin{equation} \label{linearrelation}
\log_{10} (L_\gamma) = \alpha \log_{10} (L_{RC}) + \beta, 
\end{equation}
where $L_\gamma$ is the luminosity measured by Fermi/LAT between 0.1 and 100 GeV and $L_{RC}$ is the core luminosity measured for the radio galaxy at a frequency of 5 GHz, each in units of erg/s.\footnote{By ``core'' luminosity, we refer to the emission from the central unresolved region of a given radio galaxy, typically arcseconds in extent.} We use the values of $L_{RC}$ as tabulated in Ref.~\cite{DiMauro:2013xta}.\footnote{Throughout this study, we relate the gamma-ray luminosity to the gamma-ray flux according to $F_{\gamma} = L_{\gamma} (1+z)^{2-\Gamma}/4 \pi d_L^2(z)$, where $d_L$ is the luminosity distance.} The correlation between these two quantities is not perfect, however, and there exists galaxy-to-galaxy scatter in the observed ratios of the radio and gamma-ray emission, distinct from measurement uncertainties. To account for this source-to-source variation, we take the underlying distribution of radio galaxies to exhibit values of $\log_{10} L_{\gamma}$ that are distributed around the values described by Eq.~\ref{linearrelation} with a Gaussian function. We then determine the values of the free parameters $\alpha$ and $\beta$, along with the width of the $\log_{10}$ normal distribution $\sigma$, by maximizing the negative of the log-likelihood:
\begin{equation}
-\log \like (\alpha,\beta,\sigma)=\sum_i -\log \like_i (\alpha,\beta,\sigma), 
\end{equation}
where
\begin{equation}
\like_i (\alpha,\beta,\sigma)=\int \like(L_\gamma ) \frac{1}{\sqrt{2 \pi \sigma^2}}\textnormal{exp}\left(- \frac{ \left(\log_{10}(L_\gamma ) - \alpha \log_{10} (L_{RC}) - \beta \right)^2}{\sigma^2}\right) dL_\gamma .
\end{equation}

\begin{figure}
\includegraphics[keepaspectratio,width=0.95\textwidth]{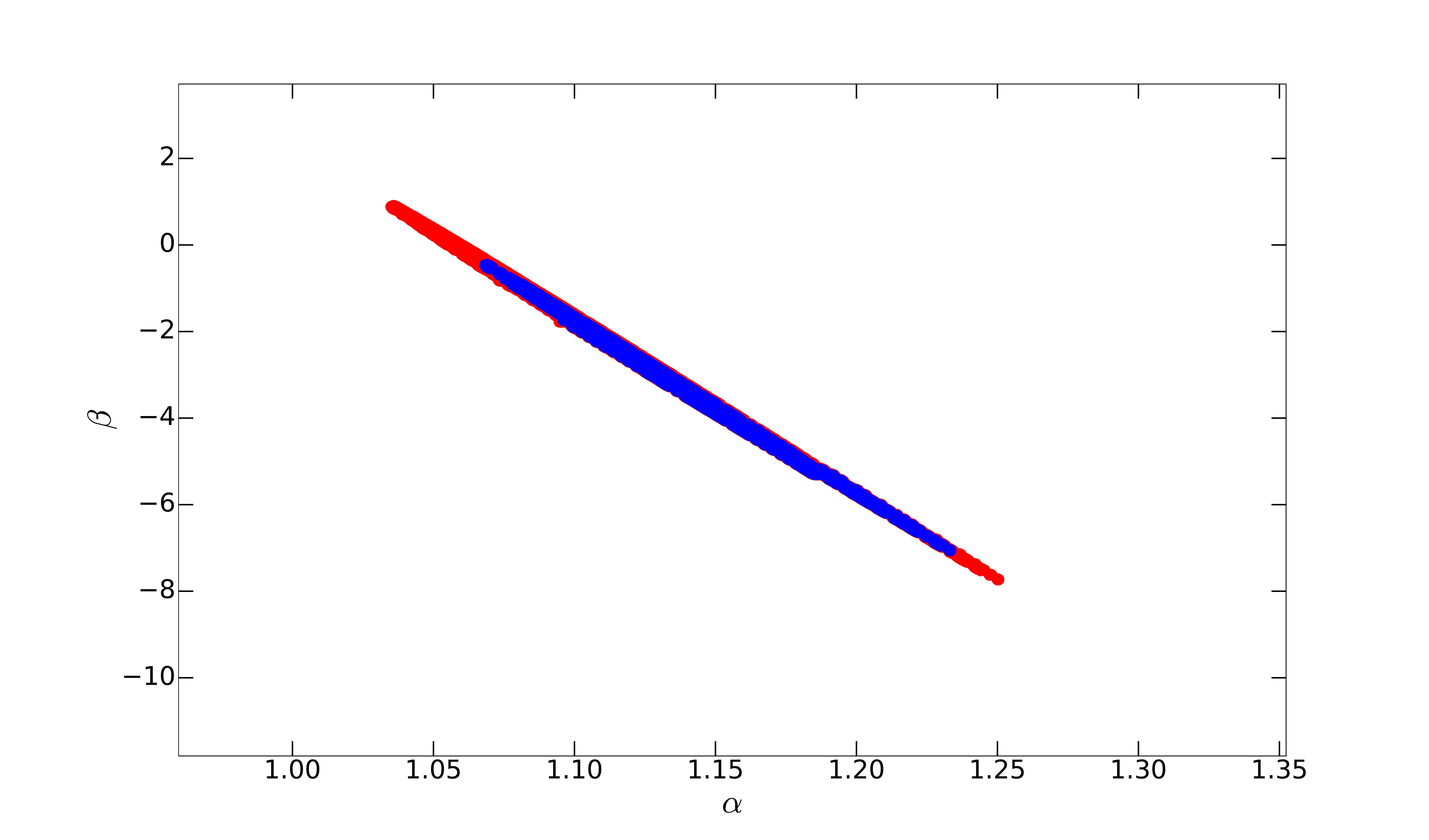}
\caption{The 1$\sigma$ (blue) and 2$\sigma$ (red) range for the parameters $\alpha$ and $\beta$ (see Eq.~\ref{linearrelation}), as determined by our fit to the 5 GHz and 0.1-100 GeV fluxes from the 51 radio galaxies in our sample.}	
\label{alphabeta}		
\end{figure}

\begin{figure}
\includegraphics[keepaspectratio,width=1.0\textwidth]{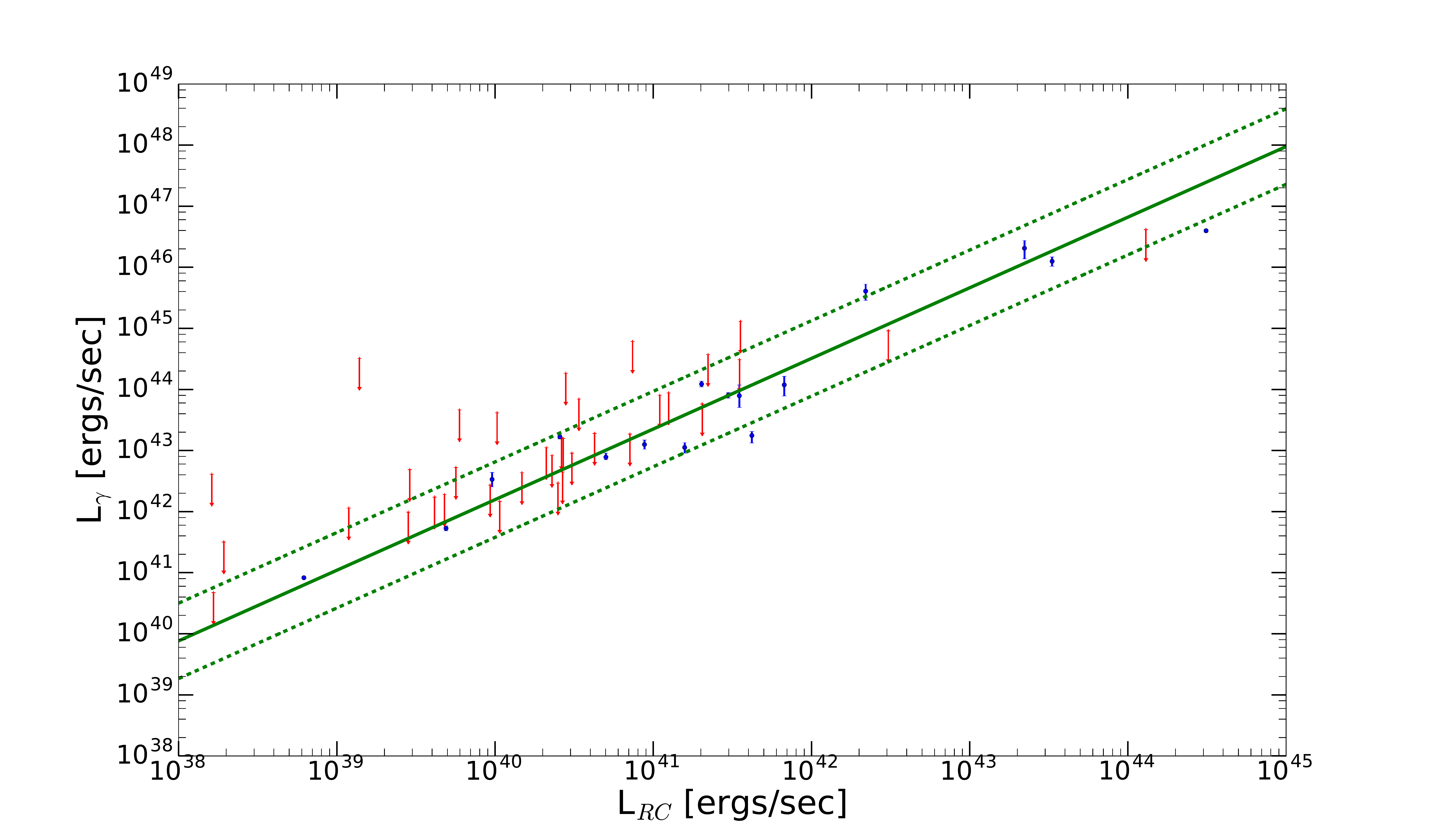}
\caption{The radio (5 GHz) and gamma-ray (0.1-100 GeV) luminosities of the 51 radio galaxies in our sample. The blue points and error bars represent sources with a significant gamma-ray detection (TS$>$25), while the red arrows denote the $2\sigma$ upper limits for those sources without such a detection.  The solid green line is the central value of the best-fit linear relationship between these quantities, and the dashed green lines enclose the 1$\sigma$ galaxy-to-galaxy variation observed among this source population.}
\label{LinearFit}
\end{figure}

We find that this exercise results in best-fit values of $\alpha=1.156$ and $\beta=-4.044$, and with galaxy-to-galaxy variation described by $\sigma=0.62$. The uncertainties in $\alpha$ and $\beta$ are correlated, and we show the error ellipses in the $\alpha$-$\beta$ plane in Fig.~\ref{alphabeta}. In Fig.~\ref{LinearFit}, we show the best fit to the $L_{RC}$ vs $L_\gamma$ relationship. The blue points denote sources with TS$>$25, while the red arrows are the $2\sigma$ upper limits for those radio galaxies with TS$<$25. The dashed lines represent the 1$\sigma$ scatter around the central value of this relationship.

Comparing our determination to the results of Di Mauro {\it et al.}~\cite{DiMauro:2013xta}, we find the following ratio of central values for the gamma-ray luminosity predicted by this correlation:
\begin{equation}
\frac{L_{\gamma, {\rm This \,  Work}}}{L_{\gamma, {\rm Di Mauro}}} \approx 1.06 \times \bigg(\frac{L_{RC}}{10^{41}\,{\rm erg}/{\rm s}}\bigg)^{0.15}.
\end{equation}
Our gamma ray-radio relationship is thus very similar to that found by the authors Ref.~\cite{DiMauro:2013xta}, although we predict slightly higher gamma-ray luminosities for the most luminous radio galaxies (those with $L_{RC} \gsim 10^{41}$ erg$/$s). We note that our procedure differs from that of Di Mauro {\it et al}. in two key respects. Firstly, Di Mauro {\it et al}. utilized only those radio galaxies detected by Fermi (with high significance) in determining the parameters of this relationship, and then used those sources without a significant detection to provide an after-the-fact consistency check. In contrast, we have included the entire sample of 51 radio galaxies in our determination, regardless of whether or not they were identified in the Fermi data, utilizing the full likelihood profile for each source. Secondly, whereas the approach of Di Mauro {\it et al}. accounts for departures from this relationship by absorbing this information into the error bars for $\alpha$ and $\beta$, our treatment allows for galaxy-to-galaxy variations in the gamma-ray flux, and we model the {\it distribution} of gamma-ray luminosities around the predicted correlation. This latter distinction leads to a higher gamma-ray flux from the sum of all unresolved radio galaxies than predicted by Di Mauro {\it et al}., as well as to smaller error bars on this flux.

\section{The Contribution to the Diffuse Gamma-Ray Background}

\begin{figure}
\includegraphics[keepaspectratio,width=0.95\textwidth]{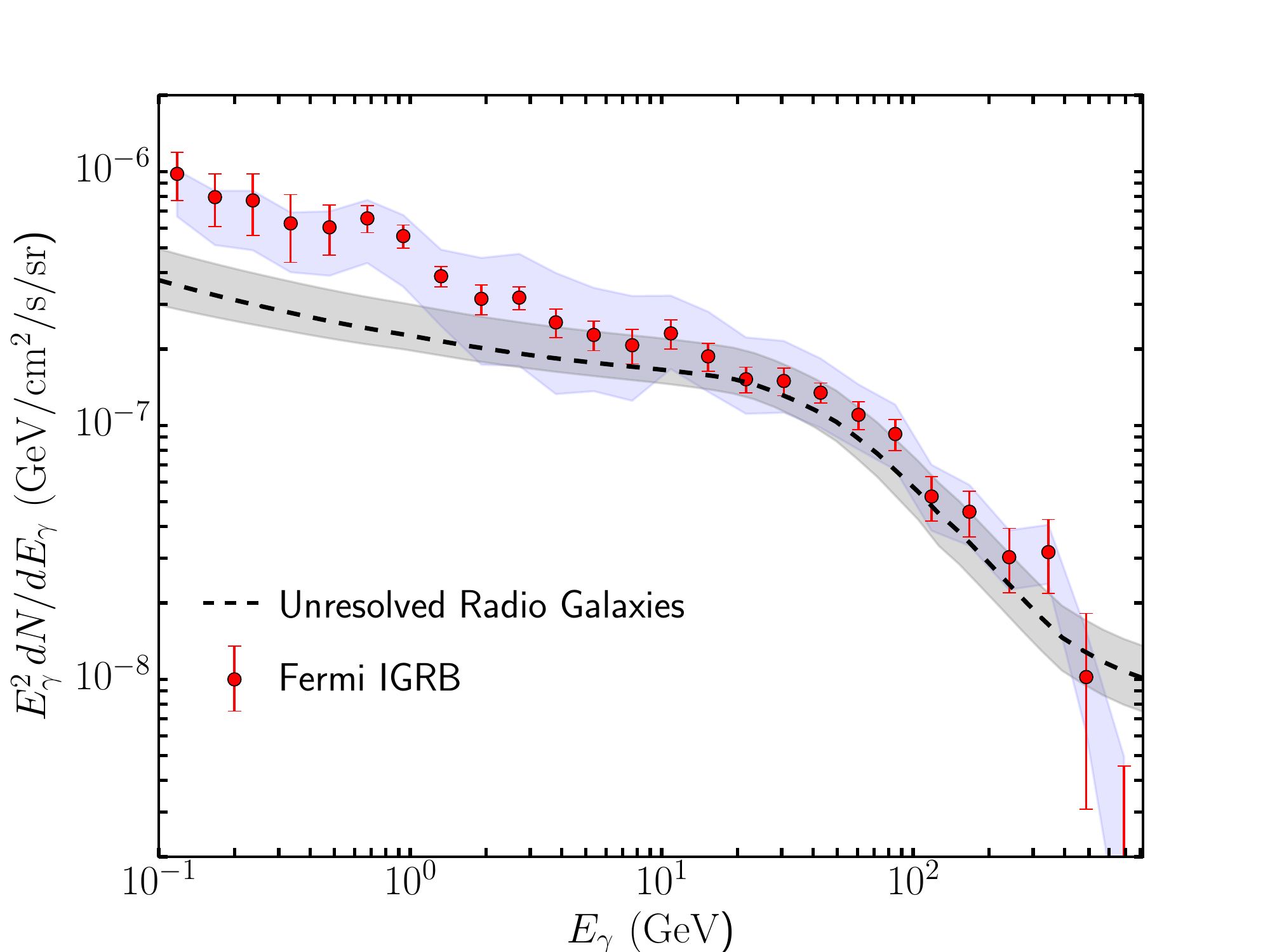}
\caption{The contribution to the diffuse gamma-ray background from unresolved radio galaxies (dashed curve), compared to the measurement of the isotropic gamma-ray background (IGRB) as reported by the Fermi Collaboration~\cite{Ackermann:2014usa}. The error bars in this figure include both the statistical uncertainty and the systematic uncertainties associated with the effective area and cosmic ray background subtraction, while the light blue shaded band reflects the systematic uncertainties associated with the modeling of the Galactic foreground emission. The grey shaded band represents the range of predicted contributions when varying the parameters describing the radio-gamma-ray correlation ($\alpha$, $\beta$) within the 1$\sigma$ range of their best-fit values. This indicates that the IGRB is dominated by emission from radio galaxies, at least at energies above approximately 1 GeV. More qualitatively, we find that $77.2^{+25.4}_{-9.4}\%$ of the photons constituting the IGRB at $E_{\gamma} > 1$ GeV originate from unresolved radio galaxies.}
\label{spec}
\end{figure}
	
\begin{figure}
\includegraphics[keepaspectratio,width=0.95\textwidth]{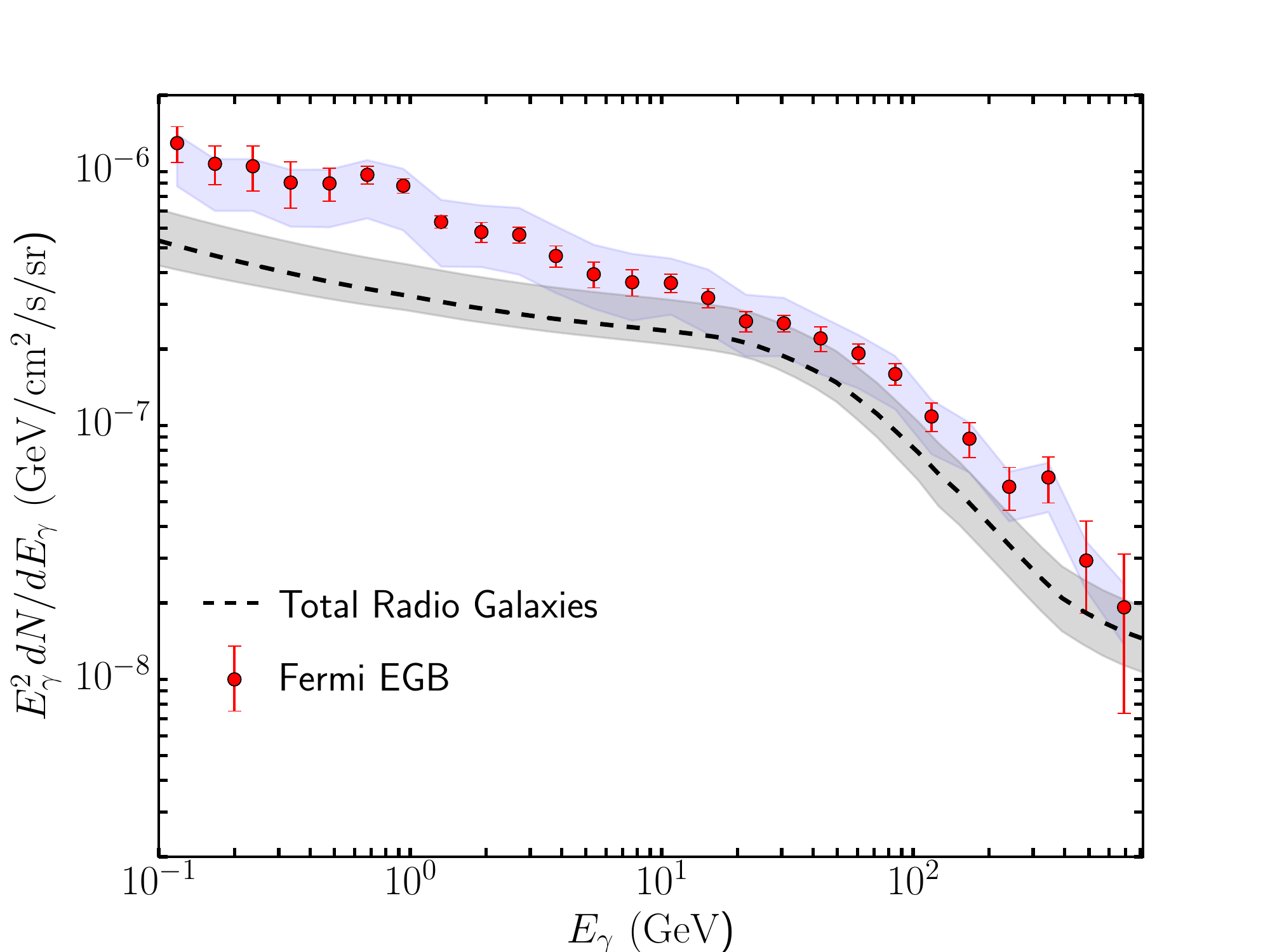}
\caption{As in Fig.~\ref{spec}, but for the Extragalactic Gamma-Ray Background (EGB)~\cite{Ackermann:2014usa}, which includes emission from both unresolved and resolved radio galaxies.}
\label{spec2}
\end{figure}

In this section, we utilize the correlation determined in Sec.~\ref{corsec} to calculate the contribution from the population of all unresolved radio galaxies to the diffuse gamma-ray background. In doing so, we closely follow the approach of Ref.~\cite{DiMauro:2013xta}.

The total gamma-ray flux from unresolved radio galaxies is given by:
\begin{eqnarray}
\label{totspec}
\frac{dF}{dE_{\gamma} \, d\Omega} &=&\int_{\Gamma_{\rm min}}^{\Gamma_{\rm max}} d\Gamma \frac{dN}{d\Gamma} \int_0^{z_{\rm max}} dz \frac{d^2V}{dz \, d\Omega} \int_{L_{\gamma,{\rm min}}}^{L_{\gamma,{\rm max}}}\frac{dF_\gamma}{dE_{\gamma}} \\
&\times& \frac{dL_\gamma}{L_\gamma \log (10)} \rho_\gamma (L_\gamma,z)(1-\omega (F_\gamma (L_\gamma ,z)))\exp (-\tau_{\gamma}(E_{\gamma},z)), \nonumber
\end{eqnarray}
where $dN/d\Gamma$ is the distribution of gamma-ray spectral indices (as determined in Sec.~\ref{gamma}), $d^2V/dz \, d\Omega$ is the co-moving volume element, and $dF_{\gamma}/dE_{\gamma}$ is the gamma-ray flux for a radio galaxy of luminosity $L_{\gamma}$ and located at redshift $z$. We integrate over the following ranges: $L_{\gamma}=10^{40}-10^{50}$ erg/s, $\Gamma=1.5-3.0$, and up to $z_{\rm max}=4$. Neglecting galaxy-to-galaxy variations in the gamma-ray to radio luminosity relationship, we have calculated the gamma-ray luminosity function of radio galaxies, $\rho_{\gamma}$, following the approach of Ref.~\cite{DiMauro:2013xta}, utilizing the total radio luminosity function of this source class as reported in Ref.~\cite{Willott:2000dh}, and the observed correlation between the total and core radio luminosities as presented in Ref.~\cite{Lara:2004ee}.\footnote{We have converted the radio luminosity function presented in Ref.~\cite{Willott:2000dh} to standard $\Lambda$CDM cosmology following the same approach described in Ref.~\cite{DiMauro:2013xta}.} To account for galaxy-to-galaxy variations, we have integrated the gamma-ray to radio correlation over the $\log_{10}$ normal distribution (of width $\sigma=0.62$) as determined and described in Sec.~\ref{corsec}. The function $\omega$ represents Fermi's point source detection efficiency, which accounts for the fact that resolved radio galaxies do not, by definition, contribute to the diffuse gamma-ray background. Although the authors of Ref.~\cite{DiMauro:2013xta} adopted an efficiency as described in Ref.~\cite{Collaboration:2010gqa} (see also the Appendix of Ref.~\cite{DiMauro:2013xta}), this is not strictly appropriate in the case of the larger Pass 8 dataset that we have utilized in this analysis. In particular, the efficiency described in Ref.~\cite{Collaboration:2010gqa} is intended to reflect the likelihood that a given source will be included in the 1FGL catalog, which contains no radio galaxies fainter than $6.6 \times 10^{-12}$ erg/cm$^2$/s. Given that our list of Fermi detected radio galaxies extends to sources as faint as $1.5 \times 10^{-12}$ erg/cm$^2$/s, we adopt as our detection efficiency the function described in Ref.~\cite{Collaboration:2010gqa}, but shifted in flux by a factor of 4.4. Lastly, the attenuation of the gamma-ray spectrum, resulting from $e^+ e^-$ pair production via scattering with the extragalactic background light, is characterized by the optical depth, $\tau_{\gamma} (E_{\gamma}, z)$. To account for this effect, we adopt the model described in Ref.~\cite{2010ApJ...712..238F}.

In Fig.~\ref{spec}, we show the predicted contribution to the diffuse gamma-ray background from unresolved radio galaxies, compared to the measurements of the isotropic gamma-ray background (IGRB) as reported by the Fermi Collaboration~\cite{Ackermann:2014usa}. The error bars in this figure include both the statistical uncertainty and the systematic uncertainties associated with the effective area and cosmic ray background subtraction. In addition, the light blue shaded band reflects the systematic uncertainties associated with the modeling of the Galactic foreground emission. The black dashed line in this figure represents our determination for the contribution to the IGRB from unresolved radio galaxies, for our best-fit values of $\alpha$ and $\beta$ (see Sec.~\ref{corsec}). The grey band around this line is the range of predictions within the 1$\sigma$ variations of these parameters. We also show in Fig.~\ref{spec2} the predicted contribution to the diffuse extragalactic gamma-ray background, which includes contributions from both unresolved and resolved radio galaxies.

This result indicates that the IGRB is dominated by the contribution from radio galaxies, at least at energies above approximately 1 GeV. More quantitatively, the fraction of the photons in the IGRB that come from unresolved radio galaxies is $77.2^{+25.4}_{-9.4}\%$ above 1 GeV.  The fraction integrated down to 0.1 GeV is significantly smaller, although still quite large, $40.0^{+13.2}_{-6.9}\%$. For comparison, the previous studies of Refs.~\cite{Inoue:2011bm} and~\cite{DiMauro:2013xta} each found best-fits in which 20-30\% of the diffuse gamma-ray background originates from unresolved radio galaxies, but with large uncertainties, extending from roughly $\sim$$10\%$ to $100\%$ of the observed flux.

In deriving this result, we assumed that the galaxy-to-galaxy variations around the (logarithm of the) gamma-ray luminosity predicted by the relationship of Eq.~\ref{linearrelation} are described by a Gaussian. Given the relatively small number of radio galaxies in our sample, however, it is possible that the true distribution is not a simple Gaussian. Most potentially important to our results are the tails of this distribution. From the distribution shown in Fig.~\ref{LinearFit}, one can see that the most significant outliers in our sample are approximately $\sim$$1.4\sigma$ away from the central value of this relationship. If we suppress all variations beyond $1.4\sigma$ ($2\sigma$) from the central value, the overall contribution to the gamma-ray background shown in Fig.~\ref{spec} is reduced by a factor of 1.5 (1.2).

\begin{figure}
\includegraphics[keepaspectratio,width=0.95\textwidth]{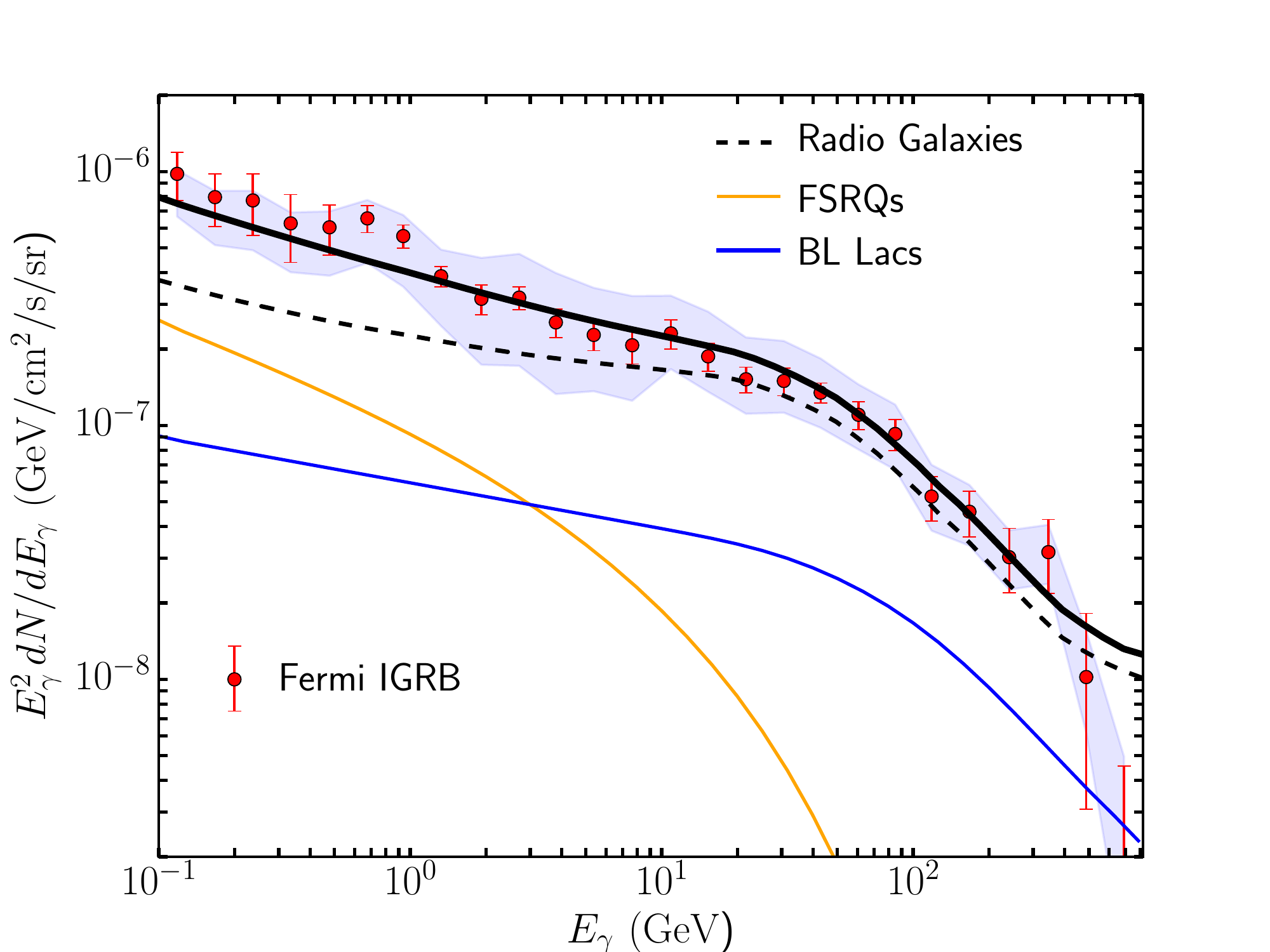}
\caption{A plausible model for the origin of the  isotropic gamma-ray background (IGRB), consisting of contributions from radio galaxies and blazars (flat spectrum radio quasars and BL Lac objects); the sum of these contributions is shown as a thick black line.}
\label{combinedspectrum}
\end{figure}

\section{Implications and Discussion}

The results described in this paper suggest that unresolved radio galaxies provide the dominant contribution to the diffuse gamma-ray background, at least at energies greater than approximately $\sim$1 GeV. This has implications not only for radio galaxies themselves, but for other classes of gamma-ray sources that may contribute to the diffuse gamma-ray background. 

In Fig.~\ref{combinedspectrum}, we show a plausible model for the major contributions to the diffuse gamma-ray background. Here, we have adopted the best-fit flux from unresolved radio galaxies as calculated in this study (black dashed), along with the contributions from unresolved flat-spectrum radio quasars (yellow solid) and BL Lac objects (blue solid), as calculated in previous studies utilizing population models (see Ref.~\cite{Cholis:2013ena}; based in turn on the results of Refs.~\cite{Ajello:2013lka,Ajello:2011zi,Collaboration:2010gqa,Cuoco:2012yf}). It is remarkable that the emission from this combination of source classes is capable of reproducing both the normalization and the spectral shape of the diffuse gamma-ray background.

The small-scale characteristics of the diffuse gamma-ray background also support a scenario similar to that depicted in Fig.~\ref{combinedspectrum}. First, the degree of small-scale anisotropy in the diffuse gamma-ray background can be used to constrain the contribution from highly luminous sources, and blazars in particular~\cite{Ackermann:2012uf,Cuoco:2012yf}. More specifically, the measured anisotropy, combined with the blazar source count distribution, suggests that blazars are responsible for $\sim$20\% of this background~\cite{Cuoco:2012yf}, consistent with that shown in Fig.~\ref{combinedspectrum}. Second, scenarios in which the diffuse gamma-ray background is dominated by emission from radio galaxies (or from starforming galaxies) are favored by the analysis of Ref.~\cite{Xia:2015wka}, which identified a significant (3-$4\sigma$) angular correlation between the diffuse gamma-ray background and the galaxies contained within the 2MASS, NVSS, and SDSS catalogs.

The results presented here leave relatively little room for contributions from other gamma-ray source classes, most notably starforming galaxies~\cite{Tamborra:2014xia,Ackermann:2012vca}, but also galaxy clusters~\cite{Zandanel:2014pva}, millisecond pulsars~\cite{Calore:2014oga,Hooper:2013nhl}, and propagating ultra-high energy cosmic rays~\cite{Taylor:2015rla,Ahlers:2011sd}. Similarly, as these results allow us to account for the origin of a larger fraction of the diffuse gamma-ray background, they will also make it possible to place more stringent constraints on any contribution that might arise from annihilating dark matter. And although the majority of the diffuse gamma-ray background does not appear to arise from dark matter annihilation products, it is plausible that a subdominant fraction of this emission could arise from such interactions. For example, if we adopt a dark matter scenario motivated by the spectrum and intensity of the Galactic Center gamma-ray excess, $m_{\rm DM} \simeq 40$ GeV, $\sigma v _{\,b\bar{b}}  \sim 10^{-26}$ cm$^3/$s~\cite{TheFermi-LAT:2015kwa,Daylan:2014rsa,Calore:2014xka,Abazajian:2014fta,Hooper:2011ti,Hooper:2010mq,Goodenough:2009gk}, one predicts that on the order of $\sim$\,$20\%$ of the diffuse gamma-ray background between $\sim$1-5 GeV should be generated through dark matter annihilations (with sizable uncertainties associated with dark matter substructure). Such analyses can probe interesting regions of dark matter parameter space~\cite{Ackermann:2015tah,DiMauro:2015tfa,Cholis:2013ena,Liu:2016ngs}, yielding limits that are competitive with those based on observations of the Galactic Center~\cite{Daylan:2014rsa,Calore:2014xka,Hooper:2012sr} or dwarf spheroidal galaxies~\cite{Drlica-Wagner:2015xua,Geringer-Sameth:2014qqa}.

Lastly, we point out that the results presented here bolster the possibility that radio galaxies (perhaps along with blazars) may be responsible for the astrophysical flux of high-energy (30 TeV-2 PeV) neutrinos reported by the IceCube Collaboration~\cite{Aartsen:2013jdh,Aartsen:2014gkd}. The authors of Refs.~\cite{Tjus:2014dna,Murase:2013rfa} have each suggested that proton-proton collisions associated with radio galaxies might be responsible for the observed neutrino flux. Intriguingly, it has also been pointed out that the same class of extragalactic sources could be responsible for the diffuse gamma-ray background and for IceCube's neutrino flux, if the responsible sources accelerate protons up to energies of $\sim$$10^{7}$-$10^{8}$ GeV with a spectral index slightly softer than two~\cite{Murase:2013rfa,Aartsen:2014njl}. The results presented in this paper provide support for a scenario in which radio galaxies -- and in particular the more numerous class of FRI radio galaxies -- are responsible for most of the 1-100 GeV diffuse gamma-ray background, as well as for the diffuse TeV-PeV neutrino flux.

\section{Summary and Conclusions}
\label{sec:conclusions}

 In this study, we have revisited the contribution from unresolved radio galaxies (sometimes called misaligned active galactic nuclei) to the diffuse gamma-ray background. Combining information from radio and gamma-ray observations, we estimate that the sum of all unresolved radio galaxies produce $77.2^{+25.4}_{-9.4}\%$ of the photons associated with the isotropic gamma-ray background above 1 GeV, as measured by the Fermi Collaboration~\cite{Ackermann:2014usa}. The blazar source count distribution, and the measured anisotropy of the diffuse background, suggests that the remaining fraction is dominated by emission from unresolved blazars (both BL Lac objects and flat-spectrum radio quasars).  Together, we find that radio galaxies and blazars can account for the normalization and spectral shape of the observed diffuse gamma-ray background.  This result also suggests that active galactic nuclei (radio galaxies and blazars) are likely to be responsible for the bulk of the high-energy neutrinos observed by IceCube.

\bigskip
\bigskip

\textbf{Acknowledgments.} DH is supported by the US Department of Energy under contract DE-FG02-13ER41958. Fermilab is operated by Fermi Research Alliance, LLC, under Contract No. DE-AC02-07CH11359 with the US Department of Energy. TL is supported by the National Aeronautics and Space Administration through Einstein Postdoctoral Fellowship Award No. PF3-140110. AL has been supported by a DOE-SCGSR Fellowship. We acknowledge the University of Chicago Research Computing Center and the Ohio Supercomputer Center for providing support for this work.

\bibliography{radio.bib}
\bibliographystyle{JHEP}

\end{document}